\documentclass[singlecolumn,         
               showpacs,            
               preprintnumbers,     
               aps,                 
               prd,          	    
               a4paper,             
               superscriptaddress,      
               nofootinbib,         
               tightenlines,        
               floats,floatfix,11pt
               ]{revtex4-2}              
\usepackage{graphicx}  
\usepackage[a4paper, total={7.0in, 9.5in}]{geometry}
\usepackage{dcolumn}   
\usepackage{bm}  
\usepackage[normalem]{ulem}
\usepackage{subcaption}

\usepackage[normalem]{ulem}
\usepackage[utf8]{inputenc}
\usepackage{amsmath,amssymb}
\usepackage{soul}
\usepackage{float}
\usepackage{makecell}
\usepackage[thinlines]{easytable}
\usepackage{array,booktabs}
\usepackage{lipsum} 
\usepackage[colorlinks=true,linkcolor=blue,citecolor=blue]{hyperref}
\usepackage{subcaption}
\usepackage{array,makecell}

\begin{document} 

\title{\texttt{CosmoDS:} A Python toolkit for constraining cosmological models via dynamical systems analysis with Cobaya}

\author{Nandan Roy}
\email{nandan.roy@mahidol.ac.th (Corresponding Author)} 
\affiliation{NAS, Centre for Theoretical Physics \& Natural Philosophy, Mahidol University,
Nakhonsawan Campus, Phayuha Khiri, Nakhonsawan 60130, Thailand}

\author{Prasanta Sahoo }
\email{prasantmath123@yahoo.com} 
\affiliation{Midnapore College (Autonomous), Midnapore, West Bengal, India, 721101}
\affiliation{NAS, Centre for Theoretical Physics \& Natural Philosophy, Mahidol University,
Nakhonsawan Campus, Phayuha Khiri, Nakhonsawan 60130, Thailand}


\begin{abstract}

We present a toolkit, \texttt{CosmoDS}, designed to study cosmological models at the background level using dynamical system analysis within the \texttt{Cobaya} framework. Dynamical system analysis is a powerful mathematical approach for studying nonlinear systems and is widely used in cosmology to investigate the stability and evolution of different cosmological models, particularly those involving dark energy. In this code, we provide a framework for constraining cosmological models using a dynamical system formulation. Most importantly, the toolkit is directly integrated with the \texttt{Cobaya} interface, allowing users to take advantage of the sophisticated statistical and inference tools already implemented in \texttt{Cobaya} for cosmological parameter estimation and model analysis.
\href{https://github.com/Nandancosmos/CosmoDS}{https://github.com/Nandancosmos/CosmoDS}.
\end{abstract}

\maketitle

\section{Introduction}
Here we present \texttt{CosmoDS}, a toolkit designed to study cosmological models at the background level using dynamical system analysis within the \texttt{Cobaya} framework. The code numerically evolves the dynamical system describing the cosmological background and computes relevant observables such as the Hubble expansion rate and cosmological distance measures required by observational likelihoods. By interfacing directly with \texttt{Cobaya}, the toolkit allows users to take advantage of its existing statistical infrastructure to perform parameter estimation and model comparison.

The discovery of the accelerated expansion of the Universe from observations of Type Ia supernovae marked a major breakthrough in modern cosmology. Subsequent measurements from cosmic microwave background (CMB) anisotropies \cite{Planck2020}, baryon acoustic oscillations (BAO)\cite{BAO2017}, and large scale structure surveys \cite{Simon:2022lde,Yuan:2022jqf,DES:2021vln,Busch:2022pcx,Wright:2025xka} have confirmed that the present Universe is dominated by a component with negative pressure, commonly referred to as dark energy. Despite the remarkable success of the $\Lambda$CDM model in describing cosmological observations, the physical origin of dark energy remains unknown and continues to be one of the most important open problems in cosmology.

Dynamical system techniques have been widely applied in the literature to study a broad class of cosmological models, including quintessence, k-essence, interacting dark energy, and modified gravity theories \cite{Sahoo:2025cvz,Sahoo:2024dgb,Thanankullaphong:2026anl,Roy_2014,PhysRevD.111.103501,Das_2019,B_hmer_2016,Wainwright_Ellis_1997,Bahamonde_2018,roy2015dynamicalsystemsanalysisvarious}. Their dynamics are commonly analyzed using the dynamical system approach, where the cosmological equations are expressed in terms of dimensionless phase-space variables, enabling the study of fixed points and stability properties of cosmological solutions. In addition to providing insights into the theoretical behavior of these models, it is also important to confront them with observational data in order to determine the allowed parameter space and assess their viability as alternatives to the standard cosmological model.

Modern cosmological parameter estimation is typically performed using Bayesian inference frameworks such as \texttt{Cobaya} \cite{Torrado:2020dgo}, which provide an efficient interface for sampling cosmological parameters and combining different observational likelihoods. In most analyses, Boltzmann codes such as \texttt{CLASS} \cite{blas2011cosmic} or \texttt{CAMB} \cite{Lewis:1999bs} are used together with \texttt{Cobaya} to compute cosmological observables and constrain model parameters. However, many cosmological models are constructed primarily to describe the late time dynamics of the Universe and are not intended to model the detailed physics of the early Universe. For such models, implementing the full Boltzmann framework in \texttt{CLASS} or \texttt{CAMB} can be unnecessarily complex. 

In these situations, it is often sufficient to study the background cosmological evolution and its observational consequences without solving the full set of perturbation equations. Nevertheless, incorporating dynamical system formulations of cosmological models into existing inference pipelines still requires additional numerical implementation and careful integration with the parameter estimation framework.

The goal of this package is to provide a flexible and extensible framework for constraining dynamical dark energy models using modern cosmological data. The modular structure of the code allows straightforward implementation of different scalar field potentials and other dynamical dark energy scenarios.
\label{sec:introduction}

\section{Example Model Implemented}
\label{sec:example_model}

As a demonstration of the framework implemented in \texttt{CosmoDS}, we consider a spatially flat universe in which the dark energy sector is described by a canonical scalar field $\phi$, commonly referred to as quintessence. In this scenario, the scalar field evolves dynamically and can drive the late time acceleration of the universe. The background cosmological dynamics are governed by the Friedmann and acceleration equations

\begin{equation}\label{rsh01}
H^2 = \frac{\kappa^2}{3} \left( \rho_m + \frac{1}{2} \dot{\phi}^2 + V(\phi) \right),
\end{equation}

\begin{equation}\label{rsh02}
\dot{H} = -\frac{\kappa^2}{2} \left( \rho_m + \dot{\phi}^2 \right).
\end{equation}

Here, $\kappa^{2}=8\pi G$, $H=\dot{a}/a$ is the Hubble parameter and $a(t)$ is the cosmological scale factor. The quantities related to matter and the scalar field are denoted by the subscripts $m$ and $\phi$, respectively, while overdots represent derivatives with respect to cosmic time.
Each component therefore satisfies the standard continuity equation $\dot{\rho_i} = -3H(\rho_i + p_i),$ where $i \in \{m,\phi\}$. For the canonical scalar field, the energy density and pressure are given by

\[
\rho_{\phi} = \frac{1}{2}\dot{\phi}^{2} + V(\phi), \qquad
p_{\phi} = \frac{1}{2}\dot{\phi}^{2} - V(\phi).
\]

The scalar field evolution is governed by the Klein–Gordon equation

\begin{equation}\label{rsh03}
\ddot{\phi}+3H\dot{\phi}+\frac{dV(\phi)}{d\phi}=0.
\end{equation}

In this example, we consider a power-law scalar field potential of the form

\begin{equation}
V(\phi) = V_{\phi 0}\,\phi^{m}.
\end{equation}

This model is implemented within the \texttt{CosmoDS} toolkit to demonstrate how scalar field dark energy models can be studied using the dynamical system framework and constrained using cosmological observations.

To analyze the phase space behavior of the cosmological system, it is convenient to introduce dimensionless variables that transform the cosmological equations into an autonomous dynamical system. We define the following variables

\begin{align}\label{rsh06}
x^{2} &= \frac{\kappa^{2}\dot{\phi}^{2}}{6H^{2}}, \qquad
y^{2} = \frac{\kappa^{2}V(\phi)}{3H^{2}} , \qquad \lambda = -\frac{1}{\kappa V(\phi)}\frac{dV(\phi)}{d\phi}.
\end{align}

Using these variables, the cosmological equations can be rewritten as an autonomous dynamical system given by

\begin{subequations}\label{eq:autonomous1}
\begin{align}
x' &= -3x + \sqrt{\frac{3}{2}}\lambda y^2 + \frac{3}{2}x\,\mathcal{U}, \label{eq:x1}\\
y' &= -\sqrt{\frac{3}{2}}\lambda xy + \frac{3}{2}y\,\mathcal{U}, \label{eq:y1}\\
\lambda' &= -\sqrt{6}\lambda^{2} x(\Gamma -1), \label{eq:lambda1}
\end{align}
\end{subequations}

where

\begin{equation}
\mathcal{U} = (1 + x^2 - y^2),
\end{equation}

 the prime denotes differentiation with respect to the number of e-folds $N = \ln a$ and $\Gamma=\frac{V\frac{d^{2}V}{d\phi^{2}}}{\left( \frac{dV}{d\phi} \right)^{2}}$.

Several important cosmological quantities can be expressed in terms of these dynamical variables. In particular, the density parameters of the scalar field and matter components are given by

\begin{equation}
\Omega_{\phi} = x^2 + y^2 , \qquad \Omega_{m} = 1 - x^2 - y^2 .
\end{equation}

The equation of state parameter of the scalar field and the effective equation of state parameter of the universe are

\begin{equation}
w_{\phi} = \frac{x^2 - y^2}{x^2 + y^2} , \qquad w_{\rm eff} = x^2 - y^2 .
\end{equation}

\begin{figure*}
    \centering
    \includegraphics[width=1\linewidth]{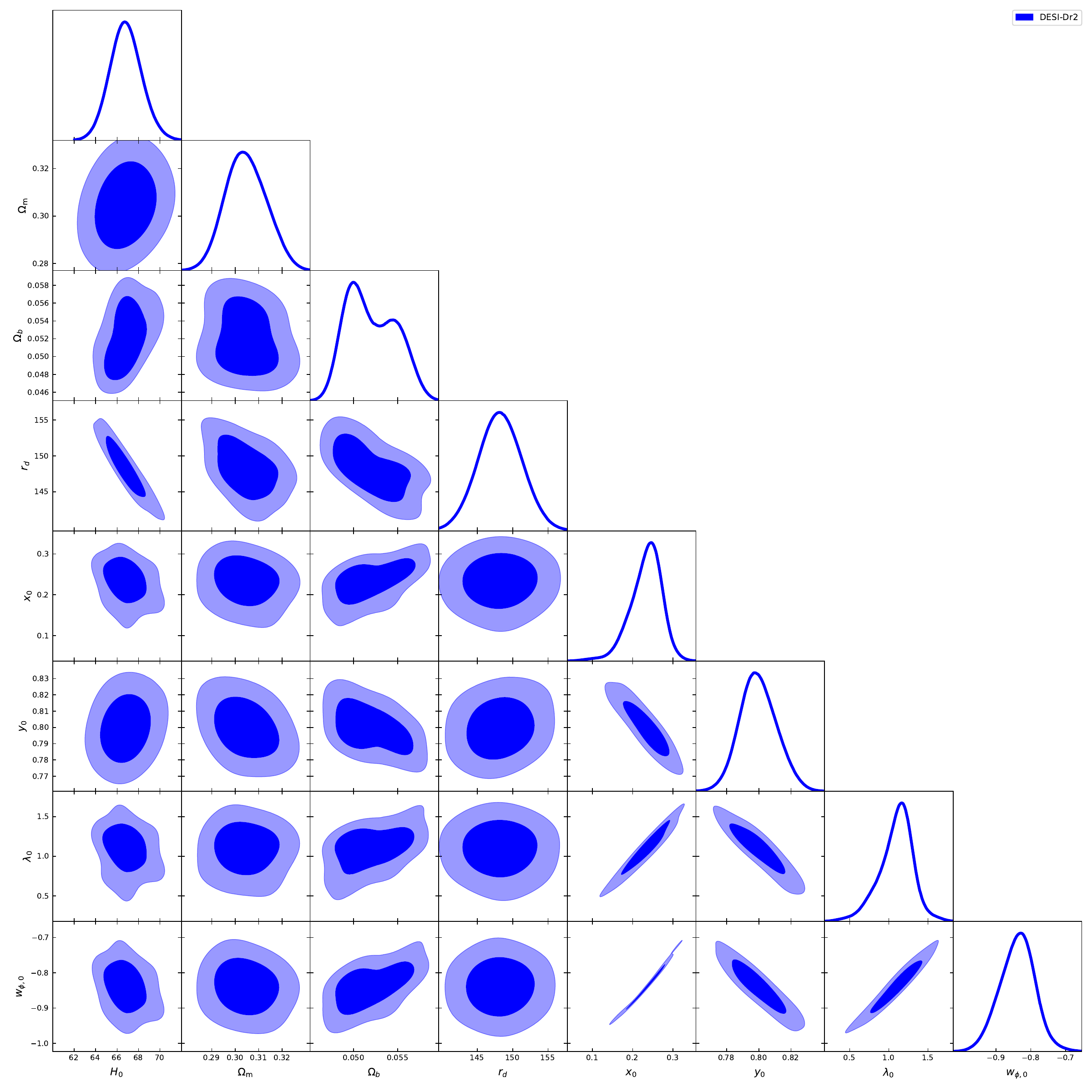}
    \caption{1D and 2D posterior distributions of the example model with $m=2$ obtained using \texttt{CosmoDS} with the combination of DESI-DR2 and DES Y5 datasets.}
    \label{fig:cosmo}
\end{figure*}

These relations allow the cosmological background evolution and the physical properties of the scalar field to be directly interpreted within the dynamical phase space framework. In the following sections, the evolution of the system can be analyzed and the model parameters can be constrained using cosmological observations within the \texttt{CosmoDS} pipeline (see Fig.\ref{fig:cosmo}).

\section{Numerical Implementation}
\label{sec:numerical}

The dynamical system described in Sec.~\ref{sec:example_model} is implemented in the \texttt{CosmoDS} toolkit to compute the cosmological background evolution. The system of autonomous equations is numerically integrated using standard ordinary differential equation (ODE) solvers available in the \texttt{SciPy} scientific computing library. In particular, the evolution equations are solved using the \texttt{solve\_ivp} integrator, which provides adaptive step size control and stable numerical performance across a wide range of cosmological parameters.

The evolution of the system is performed using the number of e-folds, $N = \ln a$, as the independent variable. This choice is commonly adopted in cosmological dynamical system analyses since it naturally tracks the expansion history of the universe and allows efficient numerical integration from the present epoch to high redshift.

For a given set of model parameters and initial conditions, the dynamical variables $(x,y,\lambda)$ are evolved simultaneously with the Hubble parameter. The numerical solution provides the background cosmological quantities required to compute observational predictions, including

\begin{itemize}
\item the Hubble expansion rate $H(z)$,
\item the luminosity distance $D_L(z)$,
\item the angular diameter distance $D_A(z)$.

\item  the angular scale of the sound horizon at the time  of decoupling ($\theta_s$).
\end{itemize}

To facilitate efficient likelihood evaluations during parameter estimation, the computed background quantities are interpolated as functions of redshift. These interpolated functions are then used to evaluate cosmological observables at arbitrary redshift values required by the observational datasets. In addition, quantities relevant for compressed CMB likelihood analyses are also computed within the code, allowing users to incorporate CMB compressed likelihoods in their parameter estimation if desired.

\section{Interface with \texttt{Cobaya}}
\label{sec:cobaya_interface}

The \texttt{CosmoDS} framework is designed to interface directly with the \texttt{Cobaya} \cite{Torrado:2020dgo} cosmological inference pipeline. In \texttt{Cobaya}, cosmological models are implemented through modular \texttt{Theory} classes that compute theoretical predictions for cosmological observables.

Within this structure, \texttt{CosmoDS} provides a custom \texttt{Theory} module that performs the following steps during each likelihood evaluation:

\begin{enumerate}
\item Read the cosmological and model parameters sampled by the inference engine.
\item Integrate the dynamical system equations describing the scalar field evolution.
\item Compute the background cosmological quantities required by observational likelihoods.
\item Provide interpolated cosmological functions to the likelihood modules.
\end{enumerate}

This design allows the dynamical system solver to function as a drop-in replacement for traditional Boltzmann solvers when only background cosmological observables are required. As a result, models that are primarily intended to describe late time cosmology can be efficiently studied without implementing the full perturbation framework typically required by Boltzmann codes.

\section{Parameter Estimation Framework}
\label{sec:parameter_estimation}

The toolkit enables Bayesian parameter estimation for dynamical dark energy models using the Markov Chain Monte Carlo (MCMC) sampling algorithms available within \texttt{Cobaya}. The framework allows users to explore the parameter space of cosmological models by combining multiple observational datasets.

The typical set of parameters sampled in the analysis includes the standard cosmological parameters such as the present day Hubble parameter $H_0$, the matter density parameter $\Omega_m$, and the baryon density parameter $\Omega_b$, along with model specific parameters describing the scalar field dynamics. In the example implementation discussed in this work, these include the initial dynamical system variables and parameters controlling the scalar field potential.

Derived cosmological quantities such as the scalar field equation of state and density parameters are computed directly from the dynamical variables during the numerical evolution. These quantities can then be used for further analysis of the cosmological behaviour of the model.

\section{Example Likelihood Analysis}
\label{sec:likelihood}

To demonstrate the capabilities of the \texttt{CosmoDS} framework, we provide example configurations that combine several commonly used cosmological datasets. These include measurements from Type Ia supernovae, baryon acoustic oscillations likelihoods.

The modular structure of \texttt{Cobaya} allows these likelihoods to be easily combined with the dynamical system solver implemented in \texttt{CosmoDS}. During the MCMC sampling process, the dynamical system is solved for each sampled parameter set, and the resulting cosmological observables are compared with the observational data.

This approach enables a consistent and efficient framework for constraining scalar field dark energy models using modern cosmological observations.

\section{Code Availability}
\label{sec:availability}

The \texttt{CosmoDS} toolkit is implemented in Python and is publicly available as an open source package. The source code, documentation, and example configurations are provided in the project repository \href{https://github.com/Nandancosmos/CosmoDS}{https://github.com/Nandancosmos/CosmoDS}.

\section{Conclusion}
\label{sec:conclusion}

In this work, we present \texttt{CosmoDS}, a toolkit designed to study cosmological models at the background level using dynamical system analysis within the \texttt{Cobaya} inference framework. The package provides a flexible and modular implementation that enables scalar-field dark energy models to be analyzed using phase-space variables while remaining fully compatible with modern Bayesian parameter estimation techniques. A key advantage of \texttt{CosmoDS} is its seamless integration with \texttt{Cobaya}, allowing users to perform sophisticated parameter estimation for cosmological models using observational data while focusing solely on the background evolution.

We expect that \texttt{CosmoDS} will serve as a useful tool for researchers investigating dynamical dark energy models and exploring alternatives to the standard $\Lambda$CDM cosmological model.

\section{Acknowledgment} We would like to thank Masroor C.~Pookkillath for useful discussions during the initial stages of developing this code.

\bibliographystyle{unsrt}
\bibliography{sample}

\end{document}